\documentclass[prb,twocolumn,showpacs,preprintnumbers]{revtex4}
\usepackage{graphics}
\usepackage{graphicx}
\usepackage{dcolumn}
\usepackage{bm}
\usepackage{epsfig}
\begin{document}
\title{Current Induced Order Parameter Dynamics:  Microscopic Theory Applied to
Co/Cu/Co spin valves}
\author{P.M. Haney$^{1}$}
\email{haney411@physics.utexas.edu}
\author{D. Waldron$^{2}$}
\email{waldron@physics.mcgill.ca}
\author{R.A. Duine$^{3}$}
\email{duine@phys.uu.nl}
\author{A. S. N\'u\~nez$^{4}$}
\email{alvaro.nunez@ucv.cl}
\author{H. Guo$^{2}$}
\email{haney411@physics.mcgill.ca}
\author{A.H. MacDonald$^{1}$}
\email{macd@physics.utexas.edu}
\homepage{http://www.ph.utexas.edu/~haney411/paulh.html}
\affiliation{$^{1}$The University of Texas at Austin, Department
of Physics, 1 University Station} \affiliation{$^{2}$Centre for
the Physics of Materials and Department of Physics, McGill
University, Montreal, PQ, H3A 2T8, Canada }
\affiliation{$^{3}$Institute for Theoretical Physics, Utrecht
University, Leuvenlaan 4, 3584 CE Utrecht, The Netherlands}
\affiliation{$^{4}$Instituto de F\'isica, PUCV Av. Brasil 2950,
        Valpara\'iso Chile}
\date{\today}
\date{\today}

\begin{abstract}
Transport currents can alter alter order parameter dynamics and
change steady states in superconductors, in ferromagnets, and in
hybrid systems.  In this article we present a scheme for fully
microscopic evaluation of order parameter dynamics that is
intended for application to nanoscale systems. The approach relies
on time-dependent mean-field-theory, on an adiabatic
approximation, and on the use of non-equilibrium Greens function
(NEGF) theory to calculate the influence of a bias voltage across
a system on its steady-state density matrix. We apply this scheme
to examine the spin-transfer torques which drive magnetization
dynamics in Co/Cu/Co spin-valve structures.  Our microscopic
torques are peaked near Co/Cu interfaces, in agreement with most
previous pictures, but suprisingly act mainly on Co transition
metal $d$-orbitals rather than on $s$-orbitals as generally
supposed.
\end{abstract}
\pacs{71.20.Lp,71.20.Be,61.50.Lt} \maketitle

\section{Introduction}
\label{sec:intro}  Transport currents can be used to alter the
order parameter dynamics of metals with broken symmetries,
including ferromagnets, antiferromagnets, superconductors, and
hybrid systems containing both magnetic and superconducting
elements. A familiar and simple example of this type of phenomena
is a circuit in which current flows between normal (N) metal
elements though a superconductor (S).  Because of Andreev
reflection of the quasiparticle current at the N/S boundary, the
Cooper pair amplitude in the
superconductor is altered.  In the presence of a normal metal
transport current, the superconducting order parameter developes a
spatial gradient in the steady state which allows the condensate
to  carry current through the superconducting element. A more
recent example is quantum Hall bilayers which develop excitonic
condensates\cite{EisensteinNature} under certain conditions and
exhibit anomalies related\cite{RossiPRL} to the influence of
transport currents on interlayer phase coherence. An important
class of related phenomena which has received a large amount of
theoretical\cite{slonc,berger,bazaliy,rossier} and
experimental\cite{tsoi1, tsoi2,SMT-exp} attention over the past
decade is centered on the influence of transport currents on
magnetization in magnetic metals, particularly ferromagnetic
nanoparticles and ferromagnets containing domain walls.  In this
case the current-related torques exerted on the magnetization can
usually be understood at least qualitatively by appealing to
conservation of total spin: the magnetization torques are
understood as the reaction counterpart to the action of the
magnet's collective exchange field on the quasiparticles, {\em
i.e.} as spin-transfer torques (STTs).  In STT physics, transport
electrons can change the magnetic state of the device, by
switching the orientation, or inducing microwave frequency
oscillations in the orientation of magnetic layers.  These
phenomona hold out the promise of applications, for example for
writing magnetically stored information. Calculations of STTs
generally proceed by computing the spatial dependence of spin
currents in a circuit and invoking conservation of spin angular
momentum to infer the torque which acts on the magnetization. This
approach has so far had mixed levels of success in predicting
important quantities like critical currents for magnetization
switching.  Most calculations to date have been based on free
electron or on semi-empirical tight-binding models.

In this paper we discuss a practical scheme for estimating the
influence of transport currents on order-paramater dynamics in
superconductors, magnetic metals, or hybrid systems.  Our
approach, which is explained in detail in
Sec.~\ref{sec:formalism}, starts by assuming a time-dependent
mean-field theory in which the electron dynamics is described by a
single-particle Hamiltonian which is uniquely defined by the
density-matrix of the system.  In practice the most flexible and
powerful choice for such a time-dependent mean-field theory will
normally be spin-density functional theory, although any similar
mean-field-like approximation is consistent with the procedures
outlined below. Hartree-Fock theory, for example, has the same
structure and has the advantage of being able to describe
non-local exchange effects which might be important under some
circumstances, for example when current is carried by a partially
occupied bonding band.
We then make an adiabatic approximation by assumming that the
time-dependence of the single-particle Hamiltonian can be ignored,
and use non-equilibrium Greens function (NEGF) theory to evaluate
the influence of a bias voltage on the system density matrix. Our
implementation of NEGF theory is explained in Sec.~\ref{sec:negf}.
The adiabatic approximation is normally appropriate since the
driving terms are proportional to transport current and are
typically weak compared to characteristic energy scales.  Since
our interest is in nanoscale systems, the neglect of inelastic
scattering in the system implied by the use of steady state NEGF
theory is normally justified. (Inelastic scattering lengths under
ambient conditions are typically $\sim 10{\rm nm}$. The approach
could be extended\cite{rembertinelastic} to include inelastic
scattering.) The influence of current on order parameter dynamics
follows from the dependence of the density-matrix on bias voltage
for any given single-particle Hamiltonian.

In the remaining sections of the paper, we apply our theory of
current-induced order parameter dynamics to a Co/Cu/Co spin-valve
system, an example of transition metal ferromagnet spintronics
which is of great practical importance and has recieved
considerable attention.
In Sec. ~\ref{sec:details} we descibe this {\em ab initio}
calculation, for which we choose a geometry similar to that
studied in Ref.~\onlinecite{katine}, in detail.  This calculation
is intended to illustrate that {\em ab initio} detail will play a
key role in designing magnetic nanostructures whose non-linear
magnetotransport properties are optimized for particular
applications.  In Sec.~\ref{sec:results}, we review the results,
and discuss some of the surprising insights that this approach
gives into the microscopic physics of spin transfer. Finally, in
Sec.~\ref{sec:conclusions}, we review our results and outline some
other interesting potential applications of this microscopic
theory of order parameter dynamics in nanostructures.

\section{Microscopic Theory of Current-Induced Order Parameter Dynamics}
\label{sec:formalism} The practical scheme for microscopic
calculation of order parameter dynamics outlined in this section
is based on a previously decribed microscopic theory of
spin-transfer torques in circuits containing ferromagnetic metal
nanoparticles.\cite{nunez} This scheme views spin-transfer torques
as a specific example of a more general class of phenomena in
which collective dynamics is altered by a bias voltage because of
the change it induces in the relationship between the
single-particle density matrix and the effective single-particle
Hamiltonian.  This microscopic view of current-induced order
parameter dynamics in magnetic metals has consequences that might
be significant in some instances, for example in circuits
containing antiferromagnetic metal elements.\cite{afm} In this
section we explain the approach, using a notation that is
convenient for the NEGF calculations we apply Co/Cu/Co spin-valves
in the body of the paper.

\subsection{Order Parameter Dynamics in an Isolated System}
When an interacting electron system is described by a
time-dependent mean-field theory the effective single-particle
Hamiltonian can be constructed at each instant in time from the
single-particle density matrix.  Historically the first example of
this type of theory is the Hartree-Fock approximation.  In modern
usage density-functional-theory is usually more accurate and
easier to implement. For example, in the local
spin-density-approximation, the Hamiltonian can be constructed
from the density and the spin-density, {\em i.e.} from elements of
the single-particle density matrix ${\bf \rho}$ that are diagonal
in a position representation. The exchange and correlation
potentials in the Hamiltonian ${\bf H}$ are explicit non-linear
functions of the density matrix: ${\bf H} = {\bf H}[{\bf \rho}]$.

We assume that a procedure for constructing the single particle
Hamiltonian from the density matrix has been chosen.  Then the
dynamics of ${\bf \rho}$ is specified by its equation of motion:
\begin{equation}
\label{eom}
\partial_t {\bf \rho} = \frac{1}{i\hbar} [{\bf H}[{\bf \rho}], {\bf \rho}] .
\end{equation}
For systems with broken symmetries, for example ferromagnets, the
order parameter is specified by a particular average of the
density  matrix so that Eq. (\ref{eom}) specifies its dynamics. In
this purely electronic equation of motion we have ignored coupling
between the electronic system and nuclear spins, phonons, or other
environmental degrees of freedom which can play a role in some
circumstances.  Such couplings are accounted for by
phenomenological additions to the right hand side of Eq.
(\ref{eom}) or to appropriate averages of this equation.

This paper focuses on the influence of transport currents produced
by a bias voltage on the collective magnetization dynamics of a
small magnetic nanoparticle.  It is helpful to consider first an
isolated system in which transport currents are absent.  We
separate both the single-particle Hamiltonian and the density
matrix into its spin-dependent and spin-independent contributions:
\begin{eqnarray}
\label{spindependence}
\rho_{i'i;s's} &=& \frac{1}{2} \big[ \rho^{(0)}_{i'i} \delta_{s's} + \vec{m}_{i'i} \cdot {\vec \tau}_{s's} \big]. \nonumber \\
{\rm H}_{i'i;s's} &=& {\rm H}^{(0)}_{i'i} \delta_{s's} -
\frac{1}{2} \vec{\Delta}_{i'i} \cdot {\vec \tau}_{s's}.
\end{eqnarray}
where $\vec \tau$ is the vector of Pauli spin matrices, $i'i$ are
orbital indices, and $s's$ are spin-indices.  The notation for the
spin-dependent part of the Hamiltonian is chosen to emphasize that
it produces a spin-splitting $\Delta$ when it is orbital
independent, as often assumed in simple toy models of a
ferromagnetic metal.  In mean field approximations, the
interaction contribution to ${\vec \Delta}$ and ${\vec m}$ are
locally related according to
\begin{equation}
\label{lsda} {\vec \Delta}=\Delta_0(n,m) \frac{{\vec{m}}}{m}
\end{equation}
 where $n$ and ${\vec m}$ are the
local charge and spin densities, respectively, and $\Delta_0$ is
some paramerization of the exchange-correlation potential. Given
this notation, it is possible to derive a useful expression for
the time-dependence of the the $\alpha$-th component of the
spin-density $\vec S$ of a chosen subsystem (SS)
\begin{eqnarray}
\dot{S}^{\alpha} &=& \mathop{\sum_{i \in SS}}_{j \notin SS}
\Big[\frac{i}{2i\hbar}  \big[{\rm H}^{(0)}_{ij} m^{\alpha}_{ji} -
m^{\alpha}_{ij} {\rm H}^{(0)}_{ji}
+ \Delta^{\alpha}_{ij} \rho^{(0)}_{ji} - \rho^{(0)}_{ij} \Delta^{\alpha}_{ji} \nonumber \big] \nonumber \\
 &+&\mathop{\sum_{i \in SS}}_j \; \frac{1}{4} \; \epsilon_{\alpha,\beta,\gamma} \; \big[ \Delta^{\beta}_{ij} m^{\gamma}_{ji} + m^{\gamma}_{ij}
\Delta^{\beta}_{ji} \big] \Big]. \label{localDsDt}
\end{eqnarray}
The first four terms on the right hand side of Eq.
(\ref{localDsDt}) represent the net spin-current into the
subsystem which has contributions from both the spin-polarization
of inter-orbital coherence and from the spin-dependence of the
inter-orbital matrix elements in the Hamiltonian.  In the final
term, contributions in which $j \notin SS$ represent an additional
spin current from spin-dependent hopping. The contributions from
$j \in SS$ describes precessional time-evolution of spins in the
subsystem under the influence of effective magnetic fields implied
by the spin-dependent terms in the Hamiltonian. We make use of
this expression below.

It follows from Eq. (\ref{localDsDt}) that the total spin of an
isolated magnetic nanoparticle satisfies
\begin{equation}
\label{npspindynamics} \hbar \vec{\dot{S}} = \frac{1}{2} {\rm
Tr}[\dot{\bf \rho} {\vec \tau}] = \frac{1}{2} {\rm Tr}[\vec{m}
\times {\vec \Delta}].
\end{equation}
If we assume that the magnetization in the nanomagnet is collinear
(or at least nearly so), the interaction contribution to the
spin-dependent part of the Hamiltonian, will be in the same
direction as the magnetization, according to Eq. (\ref{lsda}).  It
follows that electron-electron interactions, which always provide
the dominant contribution to $\vec{\Delta}$, do not directly
influence the total spin-dynamics. When the only other
spin-dependent terms in the Hamiltonain are due to an external
magnetic field which produces an orbital independent splitting
field $\vec{\Delta}^{(Z)}$, Eq. (\ref{npspindynamics}) reduces to
\begin{equation}
\label{LLeq} \hbar \vec{\dot{S}} = \vec{S} \times
\vec{\Delta}^{(Z)} .
\end{equation}
More generally, additional spin-dependent terms in the Hamiltonian
due to spin-orbit coupling and magnetic dipole-dipole interactions
produce additional contributions to the effective field so that
$\vec{\Delta}^{(Z)}$ is replaced by an effective magnetic field
$\vec{\Delta}^{(eff)}$ that depends on the total spin orientation
and includes magnetocrystalline anisotropy, shape anisotropy, and
the dissipative contribution due to the coupling between the total
spin and incoherent particle hole excitations. These effects are
normally described phenomenologically using the Landau-Liftshitz
Gilbert equation:
\begin{equation}
\label{LLGeq} \hbar \vec{\dot{S}} = \vec{S} \times
\vec{\Delta}^{(eff)} + \alpha \hbar \frac{\vec{S}}{|\vec{S}|}
\times \vec{\dot{S}}.
\end{equation}

\subsection{Influence of a Bias Voltage}
We now address corrections to the Landau-Liftshitz-Gilbert
equations which apply when a bias voltage is applied to a metallic
nanomagnet.  We first describe our strategy for a first principles
description of current-induced order parameter dynamics using a
more general language with wider applicability and then specialize
to the magnetic spin-transfer torque case.  Our approach is based
on the non-equilibrium Greens function description of
non-interacting fermions under the influence of a bias
voltage\cite{reviews} which we describe in more detail below.  In
this approach, the bias voltage is represented by placing the
system in contact with particle reservoirs with chemical
potentials $\mu_S = \epsilon_F + e V_B/2$ and $\mu_D = \epsilon_F
- e V_B/2$ in source and drain respectively.  When the system
Hamiltonian and its coupling to source and drain electrodes is
time-independent, electrons with energies inside of the transport
window $\mu_D < E < \mu_S$ solve a time-independent Schroedinger
equation with incident-from-source scattering boundary conditions.
The Schroedinger equation solution for the system plus reservoirs
system is readily constructed using Greens function techniques.
Electrons inside and outside the transport window behave very
differently.  Given a Hamiltonian, the density matrix may be
separated into a contribution from electrons inside and outside
the transport window:
\begin{equation}
\rho_{tot}[{\bf H}] = \rho_{cond}[{\bf H}] + \rho_{tr}[{\bf H}].
\label{condplustr}
\end{equation}
where we define $\rho_{cond}[{\bf H}]$ as the contribution to the
total density matrix from all states with energy below the
transport window minimum $\mu_D$, and $\rho_{tr}[{\bf H}]$ as the
contribution from states in the transport window. (The notation
$\rho_{cond}$ is intended to suggest a {\em condensate}
density-matrix since we will ultimately take an average of the
density matrix which highlights an observable associated with a
broken symmetry.) To make progress we limit our attention to
circumstances in which the order parameter and hence the effective
Hamiltonian changes slowly in time and the relevant contributions
to the density matrix are dominated by orbitals outside the
transport window. Given these approximations, which are soundly
based for magnetic systems at least,
\begin{eqnarray}
&\dot{\rho}_{tot}=i[{\bf H},\rho_{cond}] ( 1+\frac{\partial
\rho_{tr}}{\partial \rho_{cond}} )&
\label{rhodot1}\\
&\Rightarrow \dot{\rho}_{tot}\approx i[{\bf H},\rho_{cond}]&
\end{eqnarray}
Where we have noted that the second term of Eq. (\ref{rhodot1}) is
of order $eV/\epsilon_F$, which is negligibly small for metallic
systems.
The current-induced contribution to the density-matrix equation of
motion is due to the difference between ${\bf H}[\rho_{tot}]$ and
${\bf H}[\rho_{cond}]$:
\begin{equation}
\dot{\rho}_{CI}  = \frac{1}{i\hbar} [{\bf H}[\rho_{tot}]-{\bf
H}[\rho_{cond}],\rho_{cond}]. \label{currentinduced}
\end{equation}
Making the appropriate average for the total spin of a
nanoparticle we obtain the version of Eq. (\ref{npspindynamics})
which describes the current-induced contribution to the dynamics
of the total nanoparticle spin:
\begin{eqnarray}
\label{CInpspindynamics} \hbar \vec{\dot{S}}_{CI} =  \frac{1}{2}
{\rm Tr}[\vec{m}_{cond} \times [{\vec \Delta}_{tot} - {\vec
\Delta}_{cond}]]  \nonumber \\ =\frac{1}{2} {\rm
Tr}[\vec{m}_{cond} \times {\vec \Delta}_{tr} ].
\end{eqnarray}
where ${\vec \Delta}_{tot}$ and ${\vec \Delta}_{cond}$ are the
spin-dependent Hamitonians for the system arising from
$\rho_{tot}$ and $\rho_{cond}$, respectively, and $\vec
{\Delta}_{tr}$ is the difference between the two. As described
above, we consider situations in which the states in the transport
window have relatively little contribution to the order parameter
$(m_{cond} \gg m_{tr}$), and focus on exchange contributions to
${\vec \Delta}$, which are dominant, so that, according to Eq.
(\ref{lsda})
\begin{eqnarray}
{\vec \Delta}_{tot} =
\Delta_0(n,m) \frac{(\vec{m}_{cond}+\vec{m}_{tr})}{m}. \nonumber \\
\Rightarrow {\vec \Delta}_{tr} = \Delta_0(n,m)
\frac{{\vec{m}}_{tr}}{m}.
\end{eqnarray}
To evaluate the current-induced magnetization dynamics, we need to
evaulate the right hand side of Eq. (\ref{CInpspindynamics}).

We assume that we have a circuit containing a nanomagnet with
approximately collinear magnetization.  For electrons in the
transport window we can apply Eq. (\ref{localDsDt}) to obtain for
any orbital $i$ in the system
\begin{eqnarray}
0 &=&  \frac{i}{2i\hbar} \mathop{\sum_{i \in NP}}_{j \notin NP}
 \big[{\rm H}^{(0)}_{ij} m^{\alpha}_{ji} - m^{\alpha}_{ij} {\rm H}^{(0)}_{ji}
+ \Delta^{\alpha}_{ij} \rho^{(0)}_{ji} - \rho^{(0)}_{ij} \Delta^{\alpha}_{ji} \big] \nonumber \\
 &+& \; \frac{1}{2} \; ({\vec \Delta}_{cond} \times {\vec
 m}_{tr}).
\label{transportDsDt}
\end{eqnarray}
In the above equation, $\rho^{(0)}$ and ${\vec m}$ refer to the
transport contribution to the total density matrix.  In writing
the second term as a simple cross product, we have also made a
simplifying assumption that spin-dependent hopping terms are
negligible, which is valid for the systems studied here. As
explained earlier, the first group of terms on the right-hand-side
of Eq. (\ref{transportDsDt}) represents is the difference between
the spin-current into and out of the nanoparticle (NP), while the
last term represents the precession of transport electron spin in
the presence of the exchange field from the condensate. In order
to change their spin-polarization as they move from source to
drain through the nanomagnet, transport electrons must align their
spins at an angle to the exchange field. The exchange field
produced by the condensate electrons produces a torque on the
current carrying quasiparticles. Comparing with Eq.
(\ref{CInpspindynamics}), we see that this torque is equal and
opposite to the torque applied to the condensate by the transport
electrons.

In summary, to find the current induced torque on the
magnetization, we evaluate Eq. (\ref{CInpspindynamics}), or,
equivalently, the 2nd term of Eq. (\ref{transportDsDt}).  Eq.
(\ref{transportDsDt}) implies that this torque is equal to the net
spin current into or out of the nanoparticle, which is the
standard picture of spin transfer torque, and which relies on
conservation of total spin angular momentum arguments. As
emphasized in Ref.~\onlinecite{nunez}, our more general approach
is applicable to situations in which there is no conservation of
spin, and is therefore imperative in systems with spin-orbit
coupling, and in anti-ferromagnetic systems, for example.

Finally, to gain further insight into the physics of STT, we
evalute Eq. (\ref{CInpspindynamics}) locally by choosing the
subsystem to be a single atom, or single atomic orbital, as
opposed to the entire nanoparticle.  To obtain the total STT on
the nanoparticle, we sum over these individual contributions. Such
a partition of STT is strictly speaking only valid in the case
where hopping is spin-independent,\cite{spinDepExplain} which is
approximately the case for our calculations.  We have explicitly
checked that the difference in STT on the nanoparticle found is
this manner compared to evaluating Eq. (\ref{CInpspindynamics})
globally is negligible.

\section{Non-Equilibrium Greens Function Calculations}
\label{sec:negf}  To implement the ideas in
Sec.~\ref{sec:formalism}, we use non-equilibrium Green's functions
(NEGF), within a density function theory framework.  Briefly, the
Hamiltonian of the device is calculated within standard
parameterizations of LSDA,\cite{gunnarsson} generalized to include
noncollinear magnetization.  The electronic states are then
populated according the the distribution functions of the left and
right semi-infinite leads, determining a density matrix $\rho[{\bf
H}]$.  The details of the calculation procedure have been given
previously.\cite{taylor} Such formalism has been used to
successfully calculate transport properties of similar systems,
\cite{sanvito,mathon1,waldron2} and spin-dependent properties of
molecular scale systems.\cite{waldron}

Since the system under consideration here is in the metallic
regime, we work within linear response approximation for the bias
dependence $V_B$.  In this case, the expression for the current
induced torque on atom $i$ is given as
\begin{eqnarray}
\label{sdot_NEGF} \frac{{\vec {\dot S}}_{CI}} {A} =
\frac{e\mu_B}{h(2\pi)^2} \int dk_\parallel\sum_{\beta} ({\vec
\Delta}_{cond(i,\beta)} \times {\vec m}_{tr(i,\beta)})V_B.
\end{eqnarray}
where the $\beta$ is an orbital label for atom $i$, and
$k_\parallel$ refers to the transverse momentum of each
propagating state.  The transport contribution to the density
matrix is
\begin{equation}
\rho_{tr}=G^r{\rm Im}(\Sigma_L^r)G^a.
\end{equation}
Here $\Sigma^r_{L}$ is the retarded  self energy, which accounts
for the presence of left semi-infinite left lead, and $G^{r,a}$
are the retarded (advanced) advanced Green's function for the
device. The above Green's functions and self energy are evaluated
at the fermi energy, and the $\Sigma_L$ term indicates that
non-equilibrium electrons emanate from the left lead.  We have
checked explicitly that the transport density matrix satisfies Eq.
(\ref{transportDsDt})

 Within the Landauer formalism (and staying within linear response),
the current density for a bias $V_B$ is:
\begin{eqnarray}
\label{current} &&J = \frac{e^2}{h} \frac{1}{(2
\pi)^2}\sum_{\sigma} \int dk_\parallel
T_{k_\parallel,\sigma,\sigma}(\epsilon_F)V_B.
\end{eqnarray}
where the transmission coefficient is given by:
\begin{eqnarray}
T_{k_\parallel,\sigma,\sigma'} = {\rm Tr}[({\rm Im}(\Sigma^r_L)
G^r {\rm Im}(\Sigma^r_R) G^a)_{\sigma,\sigma'}].
\end{eqnarray}
The trace above refers to orbital and site. Since most experiments
are done under fixed current bias, an experimentally relevant
quantity is the total spin torque per current, given by the ratio
of Eq. (\ref{sdot_NEGF}) (summed over all atoms) to Eq.
(\ref{current}). In linear response, this is:
\begin{eqnarray}
\frac{{\vec {\dot S}}}{I} = \frac{\mu_B}{e} \frac {\int
dk_\parallel \sum_{i,\beta}({\vec \Delta}_{cond(i,\beta)} \times
{\vec m}_{tr(i,\beta)})} {\sum_\sigma \int dk_\parallel
T_{\sigma,\sigma}(\epsilon_F)}.
\end{eqnarray}

\section{Calculation details}
\label{sec:details} The system studied consists of a semi-infinite
Co lead, a Cu spacer layer with 9 atomic planes, a Co free layer
with 15 atomic planes, and a semi-infinite Cu lead.  All layers
have translational symmetry in the transverse direction, and
therefore represent the bulk realistically.  Both Co and Cu are
assumed to be in the fcc phase, and we use a lattice constant of
3.54 \AA\ throughout.  We use norm-conserving
pseudopotentials\cite{hamann} and an $s,p,d$ single-zeta basis
set.  We have found excellent agreement with established band
structure, density of states, and bulk conductivity for Co and Cu
with this basis set.  We have found that 800 $k$-points within the
Brillouin zone is sufficient for convergence of the
self-consistent density matrix. To calculate transmission
coefficients and non-equilibrium spin densities, we have used
25,600 $k$-points.  We have found that there is less than a 1 \%
difference in these quantities when using up to 32,400 $k$-points.
In calculating the spin torque contribution for different channels
in the Brillouin zone, we have found that some channels exhibit
resonance states, in which the spin density contribution at the
Fermi energy is extremely large (up to $\sim 50$ times larger than
"typical" channels).  In order to properly account for such
states' contributions to the spin density, we have isolated such
points in the Brillouin zone (if their contribution to the spin
torque/current is greater than 2 $\mu_B/e$ - such states
constitute typically at most $1\%$ of the Brillouin zone area),
and integrated over energy from 0 to .8 mV (the potential required
to reach typical critical current densities). This smooths out
their contribution to the non-equilibrium spin density.  We have
verified that points in the Brillouin zone which do not exhibit
such resonances at the Fermi energy contribute an essentially
constant spin density over this energy range, and may be safely
treated within linear response.

\begin{figure}[h]
\vskip 0.25 cm
\includegraphics[width=3.5in,angle=0]{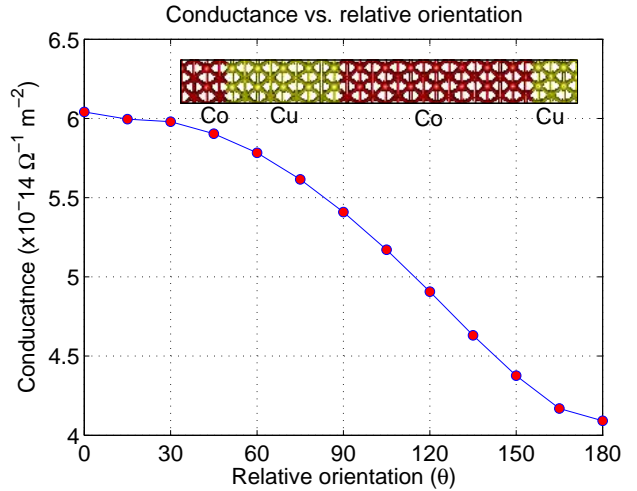}
\vskip 0.25 cm \caption{Conductivity versus relative layer
orientation.} \label{bands}
\end{figure}

Once a self-consistent collinear calculation is completed, we
initialized the various non-collinear systems by rotating the
initial spin density of the free layer to the desired angle.  We
found that non-collinear configurations do converge to
self-consistency, and ostensibly represent extrema of the total
energy. This is perhaps surprising since the exchange coupling
between the two magnetic layers implies that the energy is
extremized in parallel or anti-parallel configurations only.  We
have calculated the difference in energy between parallel and
anti-parallel alignments and have found the difference to be less
that $10^{-5}$ Hartree/orbital, thus verifying that to within the
tolerance we use for self-consistency (a maximum change of
$10^{-5}$ Hartree in the Hamiltonian between iterations of the
self-consistent cycle), a noncollinear configuration can be a
self-consistent solution. Our value for the change in energy
between parallel and anti-parallel (or the exchange energy) is
consistent with previously found values.\cite{mathon2}

\section{Results}
\label{sec:results} Fig. 1 shows the calculated conductance per
area versus angle. From the conductivity in the parallel and
anti-parallel alignment, a GMR ratio of 48\% is obtained. The
absolute value of the conductance is consistent with previous
calculations of similar systems.\cite{sanvito,mathon1} The
conductance at $\theta=90^{\circ}$ is not the mean between the
conductance of the parallel and anti-parallel state, which
indicates that the behavior can not be captured by a simple
rotation of the collinear transmission coefficients in spin-space,
but also includes higher order effects such as multiple
spin-dependent scattering between the two layers.

Fig. 2 shows the spin torques as a function of the layers'
relative orientation. An important distinction is that between
in-plane torques, which are non-energy conserving and present only
in non-equilibrium cases, and out-of-plane torques, which are the
result of itinerant electron exchange, and are responsible for
RKKY-like interactions between layers.  (In the following we refer
to the out-of-plane torques as STT($\theta$).)  According to Fig.
2, the ratio of these two torques varies with angle, but we find
it to be consistently above $10\%$, which can have important
implications for the behavior and stability of the magnetization
dynamics. \cite{edwards} Interestingly, we also find that the
out-of-plane torque undergoes a sign change near
$\theta=180^\circ$.  Previous studies have found a difference in
sign between these torques under certain conditions,
\cite{edwards} but the relative sign is usually constant over all
angles.  The presence of an angle dependent relative sign of the
torques may also have interesting consequences for the dynamics of
the magnetization.

\begin{figure}[h]
\vskip 0.25 cm
\includegraphics[width=3.5in,angle=0]    {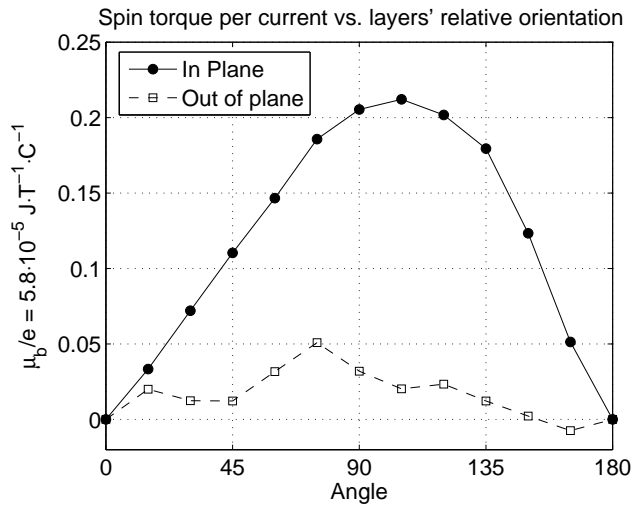}
\vskip 0.25 cm \caption{Torque per current versus relative layer
orientation.} \label{bands}
\end{figure}

The angular dependence of the conductance is partially responsible
for the departure from perfect sine behavior in STT$(\theta)$.
This departure is encapsulated in $g(\theta)$, defined as
STT$(\theta)/({\rm I \cdot sin}(\theta))$, which has been
calculated for simpler models,\cite{slonc} and is an extremely
important parameter in spin torque physics.  It represents the
amount of spin torque delivered per electron in the current flow,
and is therefore a measure of the efficiency of spin transfer
torque. Fig. 3 shows our calculated $g(\theta)$, and, for
comparison, that found in the original calulation of Sloncewski.
Note that the y-axis for the two curves are different by a factor
of 2, so that at small angle, our calculated result is more than a
factor of 2 smaller than that of Slonczewski.  That is to be
expected, as Slonczewski's model calculation considered the
limiting case of pure spin filters, where minority spins are
completely reflected, and majority spin completely transmitted
through the magnetic interfaces.  In this sense the Slonczewski
can be considered to give the most optimistic spin transfer
efficiency. Interestingly our result shows a smaller difference in
efficiency between $\theta=0$ and $\theta=\pi$ as compared to
Slonczewski, which is more consistent with experimental
data.\cite{grollier}

\begin{figure}
\vskip 0.25 cm
\includegraphics[width=3.5in,angle=0]{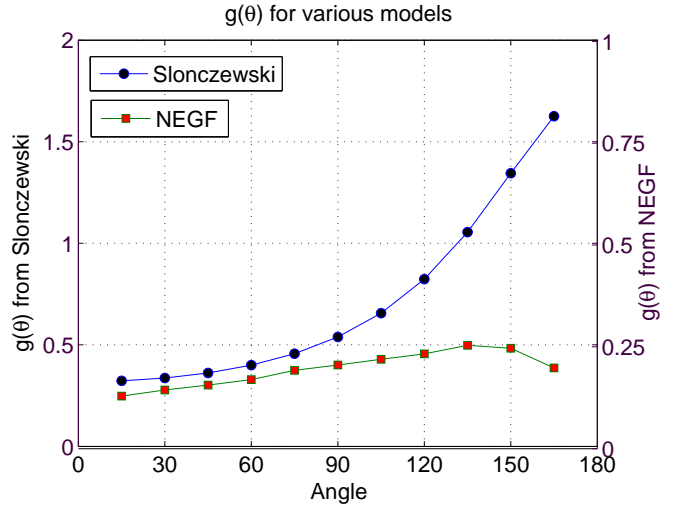}
\vskip 0.25 cm \caption{Angular prefactor of spin transfer torque
$g(\theta)$.  Note the Slonczewski scale differs from our
calculated $g(\theta)$ scale by a factor of 2.} \label{bands}
\end{figure}

One can make further comparison to experiment in finding the
amount of torque delivered per current. This quantity can be
extracted from experimental data by finding the slope of the
linear relation of critical current versus applied field.  The
experiment in Ref.~\onlinecite{katine} gives a slope of .29 mA/.1
T.  The resulting spin torque per current (or spin torque
efficiency $g(\theta)$ at $\theta=0$) is given by $\frac{\alpha
\gamma (.1T)M_s V} {.29 {\rm mA}}=.35 \mu_B/e$. Here $\alpha$ is
the bulk magnetic damping of Co, assumed to be .007, $M_s$ is the
bulk magnetization of Co, and $V$ is the volume of the free layer.
The value obtained for the efficiency in this way is nearly
identical to that determined in point contact
experiments.\cite{pufall} This contrasts with our calculated
$g(\theta)$ of .11.
Our calculated efficiency is smaller than experimental by about a
factor of 3. Remarkably, even Sloncewski's efficiency is smaller
than that seen in experiment by 50\% (assuming polarization of Co
to be .4). Some diffusive models of spin transfer typically
predict much smaller efficiencies, with values for Co-Cu
structures of about 1\%.\cite{waintal}
At first glance then, it is a puzzle as to why the efficiencies
seen experimentally are so high. One potentially important
consideration is that the deduced experimental efficiencies are
not directly measured, but are rather model dependent on the
mechanism for switching (usually simple coherent switching is
assumed), and depend on parameters such as the magnetic damping,
whose values and details are not well known.

Fig. 4 shows the layer resolved spin torques.  As expected, the
torques show a generally oscillatory decay, which is understood as
the result of averaging over many transverse channels' oscillatory
contributions.\cite{stiles}  The length of the free layer is 2.67
nm, thus our result that the transverse spin density is close to
complete decay is consistent with other approaches, in which the
decay length of transverse spin in Co is found to be
3nm.\cite{zhangLevy}

We estimate the propensity (or lack thereof) for a spin wave or
noncollinear structure in the free layer due to the non-uniform
torques by comparing the magnitude of these torques (or effective
fields) and the exchange field an atom experiences from its
neighbors.  The effective Heisenberg nearest neighbor coupling
constant for bulk Co has been computed to be 1.085 mRy.
\cite{pajda}  The resulting exchange constant between planes
(considering only nearest neighbor interactions) is then 4.34 mRy.
This corresponds to a field of about $10^3 T$.  For a current of 1
mA, the figure shows that typical non-equilibrium exchange fields
to be on the order of .0075 T.  The resulting deflections of the
spin are much less than $1^\circ$. This slight departure from
collinearity indicates that the surface-effect aspect of spin
transfer torques (more generally, non-uniform torques) are not
manifest in thin layers.

\begin{figure}[h]
\vskip 0.25 cm
\includegraphics[width=3.5in,angle=0]{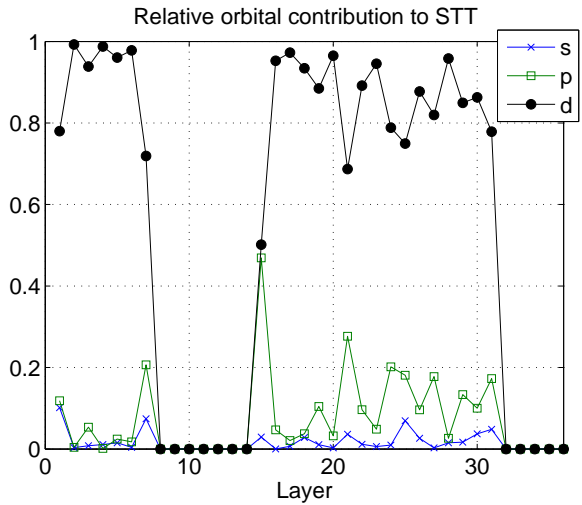}
\vskip 0.25 cm \caption{Breakdown of relative orbital
contributions to spin torque in the $z$-direction, for the
$\theta=90^\circ$ case. } \label{bands}
\end{figure}

Fig. 5 shows the relative contributions to the spin transfer from
the $s$, $p$, and $d$ orbitals.  Such a partition ignores
hybridization between orbitals, but one can nevertheless extract
meaningful results from such a division. \cite{sanvito} We find
the $d$-orbitals' contribution is dominant. This is due to the
substantial amount of $d$-electron conduction in the system, and
more importantly to the relatively larger spin-dependent
exchange-correlation potential of the $d$-electrons.  A commonly
used model for spin torque calculations in multilayers or in
domain walls is the s-d model, in which the $d$-electrons are
responsible for the exchange field, while the $s$-electrons are
responsible for the current. Spin torque on the $d$-electrons'
magnetization is a result of the interaction of these two
subsystems.  However, in the ballistic limit for 3$d$ transition
metals, our results show that it is largely the interaction
between equilibrium and non-equilibrium $d$-electrons that is
responsible for spin torque physics.

\begin{figure*}[h!]
\vskip 0.2 cm
\includegraphics[width=7in]{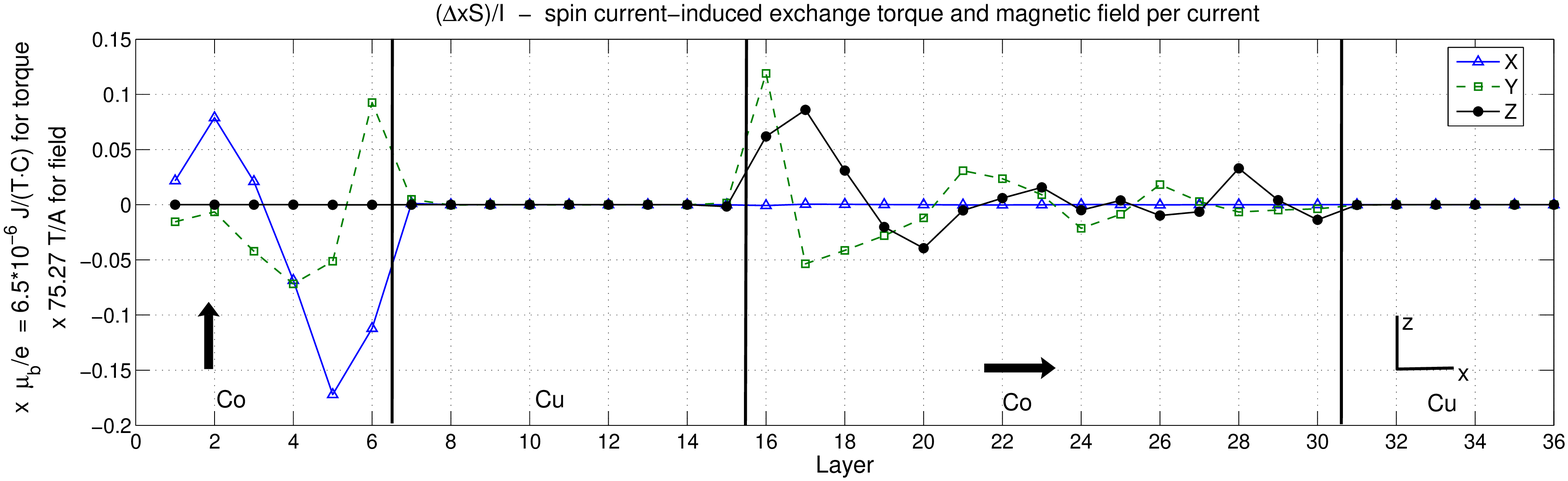}
\vskip 0.2 cm \caption{Calculated layer resolved torque per
current density and spin transfer induced change field per A. The
(X,Y,Z) directions in the legend refer to the spin torques, while
the direction of the effec } \label{bands}
\end{figure*}

\section{summary and conclusions}
\label{sec:conclusions}  In this work, we have formulated a
general scheme for calculating the dynamics of an order parameter
in the presence non-equilibrium, current carrying quasiparticles.
This scheme is applied to find the current-induced torques that
are present in magnetic spin valve structures under bias.  For the
specific system studied, we have found the STT to be localized in
the surface planes of the free magnetic layer, and have found the
out-of-plane torque to be a substantial fraction of the in-plane
torque.  The magnitude of the torque is appreciably smaller than
that deduced from experiment, although we note that experimental
measures of the absolute spin torque are model-dependent.  In
addition, we have found the STT to be due mostly to the
interaction between non-equilibrium $d$-electrons with the
exchange field, which is itself also due mostly to the spin
splitting of the $d$-band electrons.

The formalism presented here is more general than that typically
used in calculations of current induced torques in magnetic
materials.  In particular, since it does not rely on conservation
of angular momentum, it may be applied to system with spin-orbit
coupling, or to antiferromagnetic systems.  And the calculation of
current induced torques on an atomic, or even atomic orbital scale
resolution allows for the study these effects in molecular scale
systems.  More generally, it may be applied to any mean-field
system with an order parameter - for example superconductors, or
ferromagnet-superconductor hybrids.

\acknowledgments The authors acknowledge helpful interactions with
Olle Heinonen and Maxim Tsoi. This work was supported by the
National Science Foundation under grant DMR-0606489, by a grant
from Seagate Corporation, and by the Welch Foundation.  ASN was
partially funded by Proyecto MECESUP FSM0204.
Computational support was provided by the Texas Advanced Computing
Center.


\begin{thebibliography}{100}
\bibitem{spinDepExplain}
The relation between partitioning STT by orbital and
spin-dependent hopping can be understood in the following way:
When summing over orbital spin fluxes, normally intra-orbital spin
currents cancel, so that such spurious contributions do not
contribute to the total spin flux into/out of the atom.  If there
is spin-dependent hopping, such contributions do not cancel.  If
spin-dependent hopping is weak, however, these spurious
contributions are negligible.
\end{thebibliography}
\end{document}